\documentclass{article}
\usepackage[T1]{fontenc}
\usepackage[utf8]{inputenc}
\usepackage{ismir} 
\usepackage{amsmath, cite,url}
\usepackage{graphicx}
\usepackage{color}
\usepackage[bookmarks=false]{hyperref}

\title{Melodic and Metrical Elements of Expressiveness in Hindustani Vocal Music }

\multauthor
{Yash Bhake \hspace{1cm} Ankit Anand\hspace{1cm} Preeti Rao} {
Department of Electrical Engineering, Indian Institute of Technology Bombay, India\\
\texttt{yash.bhake@iitb.ac.in, ankit0.anand0@gmail.com, prao@ee.iitb.ac.in}
}

\def\authorname{Y. Bhake, A. Anand, and P. Rao}

\begin{document}

\maketitle

\begin{abstract}
This paper presents an attempt to study the aesthetics of North Indian \textit{Khayal} music with reference to the flexibility exercised by artists in performing popular compositions. We study expressive timing and pitch variations of the given lyrical content within and across performances and propose computational representations that can discriminate between different performances of the same song in terms of expression. We present the necessary audio processing and annotation procedures, and discuss our observations and insights from the analysis of a dataset of two songs in two \textit{ragas} each rendered by ten prominent artists.

\end{abstract}

\section{Introduction}
\label{sec:introduction}
A \textit{Khayal} performance consists of a composition (the chosen \textit{bandish} or song with lyrics) that also forms the base for improvisation \cite{Wade}. That is, while the first occurrence of a \textit{bandish} line is typically rendered in its canonical form from memory, the accomplished artist renders successive repetitions of the line with pitch and timing variations in specific ways to create an aesthetic experience appreciated by listeners trained in the genre. A motivation for this work is that suitable computational representations can help obtain insights or verify hypotheses regarding this performance practice. They can also inform tools to generate realistic audio for artistic and educational contexts \cite{cancino2018computational}.

Performance expression, according to Juslin \cite{juslin2000cue}, concerns "the small and large variations in timing, dynamics, timbre, and pitch that form the microstructure of a performance and differentiate it from another performance of the same music". 
While the structure of the songs themselves carries specific emotion – this certainly holds for \textit{bandish} given the \textit{raga} that underlies the melody, and the semantics of the lyrics, - expressive variations contribute to the emotional impact of the performance. We present a review of some of the work of musicologists who have studied \textit{Khayal} music from the viewpoint of composition and improvisation. 

In a scholarly work based on several vocal performances of a single \textit{bandish} by different artists, Morris \cite{Morris-AD} used his transcriptions of the recorded music (well aware of the ambiguities inherent in the manual process that involves discrete symbols) to compare an artist's rendition with their self-declared notation for the composition. He noted that some portions of \textit{bandish} were reproduced identically, others were more flexibly produced, and proceeded to investigate the extent and nature of the differences.
Another example of such work is the analysis of a single live performance by van der Meer \cite{Meer2014book}, focusing on \textit{raga}-related vocal pitch inflections in the \textit{raga} improvisation section (\textit{alap}) that triggered emotion as detected from the audience's spontaneous audible reactions. He relates these expressive moments to the occurrence/execution of specific ornaments. Finally, a close parallel of our problem is that of tracking \textit{sangatis}, which are lineage-dependent variations of the same lyric lines in Carnatic compositions \cite{sankaran2015automatic}.

More recently, the similar questions using slightly larger, curated data sets have been facilitated by the availability of computational methods. Previous computational studies have examined the variability of \textit{raga} phrases in the course of \textit{Khayal} improvisation with respect to underlying tempo and metrical location \cite{ganguli2021}. The context of singing \textit{bandish} lines, on the other hand, entails additional constraints given the special importance accorded to the lyrics in the context of \textit{bandish}. Recently,  a small comparative study of two performances of the same \textit{bandish}, with reference to temporal deviations of note events from the canonical metrical positions \cite{WIMAGA}, served as a preliminary validation for methodology using automatically detected onsets for the audio-level and text-level alignment of manual and reference syllable-level transcription. Their visual representation of the distribution, across each performance audio, of the timing offset corresponding to a specific metrical position showed the musicologically anticipated behaviour, viz. reduced deviation closer to the \textit{sam} or downbeat of the rhythm cycle. It also helped visualise the difference between the two performances in terms of the use of timing expressiveness. In contrast, the present work (i) uses a significantly larger dataset for a timing expressiveness study while also adding new attributes; (ii) includes a new study of pitch-based expressiveness; (iii) proposes computational measures for artist-level expression on each of the studied dimensions and validates their potential for discriminating performances based on expressiveness.

In the next section we provide some of the music background needed to appreciate the specific questions we set up for our work and our overall approach. Following this, we present our dataset, audio and text processing methods and the devised computational measures. Finally, we discuss our observations and attempts to draw insights.

\section{Background}


\textit{Bandish} are lyrical compositions that have been handed down across generations by oral transmission and serve as a sort of dictionary for the associated \textit{raga} in terms of the overall pitch movement and characteristic phrases \cite{srao-hcm}. The two verses of the \textit{bandish}, termed \textit{sthayi} and \textit{antara}, typically sung early on in any concert, are defined by their lyrics and tune. 
In the early 20th century, music scholar Pt. V.N. Bhatkhande launched on a mission to collect and notate traditional compositions from across the country. He devised his own ways of notating the melodies that so far lay within the framework of oral transmission and largely within hereditary musician families. 
The notation, considered a "schematic" form, requires an understanding of the intricacies of the \textit{raga} and \textit{tala} system in order to interpret in performance.
Now widely respected, his monumental work helped preserve the traditional compositions for posterity \cite{Bhatkhande}. Although commonly featured in learning contexts, it is held 
that the notation corresponding to a given performed \textit{bandish} 
is dependent to an extent on the lineage (\textit{gharana}) of the artist. Text differences exist as well, but these are mainly due to spelling and dialects or verb forms, while maintaining the recognizability of the lyrics.

\begin{figure}[ht]
    \centering
    \includegraphics[width=1\linewidth]{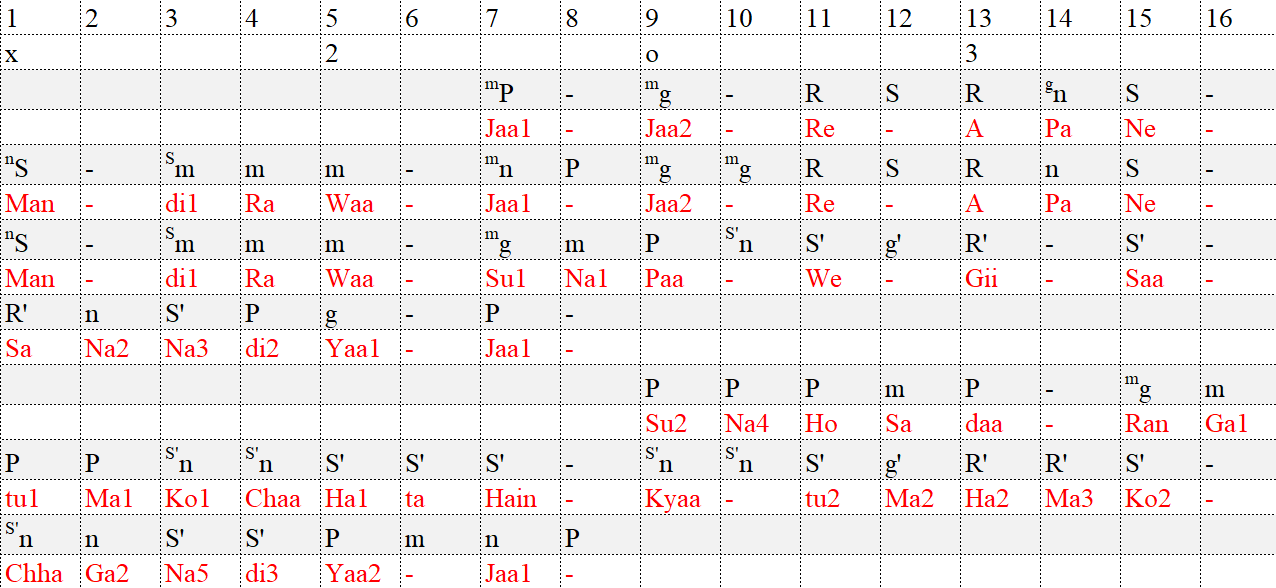}
    \caption{The composition \textit{Ja Ja Re} from Bhatkhande \cite{Bhatkhande} as a machine-readable CSV file for the two verses of 2 lines each.}
    \label{fig:Figure 1}
\end{figure}

Bhatkhande's book \cite{Bhatkhande} typically provides a single notation sequence for every unique line of the song with a \textit{raga} \textit{swar} (note), or a short sequence of \textit{swar}, assigned to every syllable of the lyrics. Figure~\ref{fig:Figure 1} shows the canonical notation of a \textit{bandish}, as available

in the book and the created machine-readable format, retaining the note label (both pitch and lyric syllable) and timing information as indicated by the labels of the 16-beat \textit{tala} cycle. In this study, all \textit{bandish} are set in \textit{teentaal} (16-beat cycle), with salient \textit{matra} (beats) like the \textit{sam} (downbeat, 'x') and \textit{khali} (9\textsuperscript{th} beat, 'o', coinciding with the start of the second half cycle) explicitly marked. The top row indicates beat number, the row below it indicates \textit{vibhag} symbols, marking the start of each quarter cycle.  Each line of lyrics of the \textit{bandish} spans 16 beats, with each beat containing either a syllable (or, seldom, multiple syllables), a rest (empty string), or a continuation of the previous note (‘s’). Each cycle is represented using two aligned rows: 
(1) Notation row (black), which displays the main note (\textit{sargam}) with any ornamentation indicated as a superscript preceding the main note (such as a grace note or a glide); and 
(2) Lyrics row (red), which contains the corresponding \textit{bandish} lyric syllables aligned to the beat positions and associated notes.


In performance, singers typically repeat each line (and sometimes its component phrases) several times with variations introduced after the first utterance is rendered in the canonical form. The successive repetitions are characterised by a certain fluidity in the notes and timing of where a given syllable falls \cite{Morris-AD}. 
These variations are normally extempore in concert settings but could show similarities across the artists from the same musical lineage (\textit{gharana}).
Viewing these as expressive variations of the \textit{bandish} line, as specified by its lyrics, we investigate the range and nature of the expressive gestures as a function of the specific word (via its component syllables) across its multiple occurrences in the singing. That is, we hope to obtain insights about the structural moments of the song where the expressive gestures are added and precisely which acoustic parameters are varied to realise this.  


\section{Dataset and processing}\label{sec:dataset}
Our dataset\footnote{More details appear in the \href{https://www.notion.so/ISMIR-2025-1c443a74812b801cab13c9946b0f83ce?pvs=4}{supplementary material}} is designed to serve the objectives of our study, namely, to observe the variations across repeated utterances of a given \textit{bandish} line across performances of the same and different artists. We consider two specific popular compositions, each with multiple performances by several prominent artists in the genre, obtained from across commercial and free internet sources. The sung \textit{bandish} lines are manually segmented from full concerts (which also typically include several other sections, including free improvisation in the chosen \textit{raga}). The details of the dataset used in this study appear in Table~\ref{tab:datasummary}. An immediate observation is the skewed distribution across the 4 \textit{bandish} lines with Line 1 being sung far more frequently than any other. This is common in the concert setting, given that the \textit{bandish} line 1 is known as the \textit{mukhda} or refrain of the song, analogous to the \textit{pallavi} in Carnatic style of Indian classical music.

We carry out vocals separation on the concert audio segments to eliminate the accompanying instruments. 
We use a pretrained model (OpenAI/Whisper) for speech to text conversion with a prompt comprising the words of the song lyrics in Devanagiri script, together with expected pronunciation variations. The obtained word sequence is then aligned at the phone level to the audio using forced alignment with a Kaldi TDNN Hindi speech trained acoustic model \cite{povey2011kaldi}. Next, the sequence of phones is segmented into the syllables of the lyrics words. Each resulting syllable's onset instant is then identified as the frame corresponding to the consonant-vowel (CV) transition. 

\begin{figure}[ht]
    \centering
    \includegraphics[width=1\linewidth]{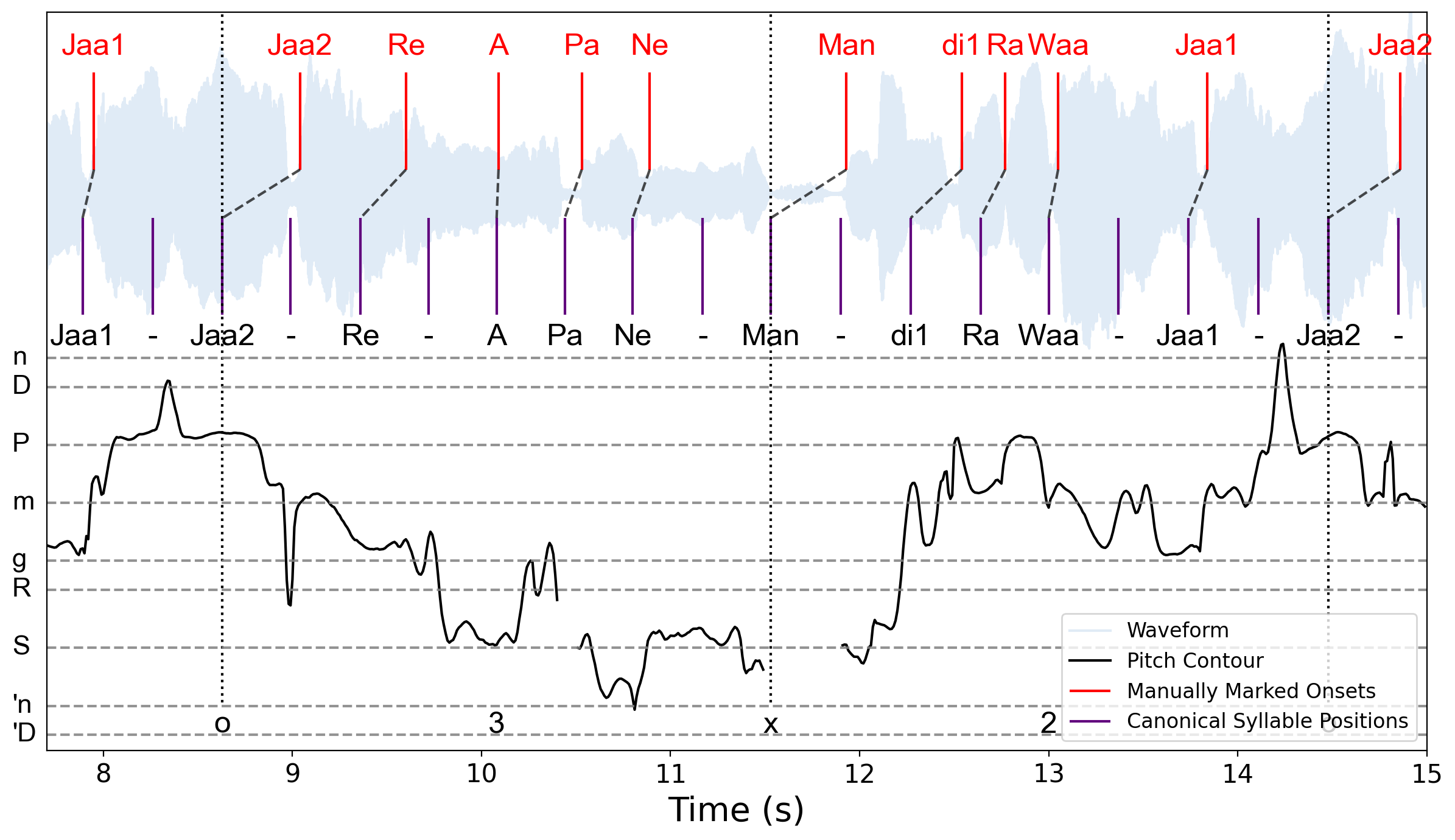}
    \caption{Pitch contour (bottom) and sung syllable alignment (red) with canonical beat positions of the same syllable (black) for an excerpt of  \textit{Ja Ja Re} by ABD. }
    \label{fig:Figure 2}
\end{figure}

The resulting alignments are checked and manually corrected for the occasional errors that arise mainly due to the presence of long vowels in singing, poor enunciation at times, as well as the occurrence of significant pitch inflections. 
We observe the manually corrected onset locations- i.e. the syllable onsets as realised by the artist- marked at the top of the waveform in the example of one full \textit{tala} cycle of the \textit{bandish} in Figure~\ref{fig:Figure 2}. To obtain the beat locations, we annotate the tabla stroke onsets using the source-separated accompaniment file; we manually mark the salient beats (downbeat 'x' \textit{sam} and 9\textsuperscript{th} beat 'o' \textit{khali}) in 1 cycle. Assuming a consistent local tempo, we divide each half cycle (interval between one downbeat and the adjacent t\textsuperscript{th} beat into 8 equal parts, thereby obtaining all the estimated beat instants across the rendition. 

Next, we mark the canonical locations of the matching syllables, positioning the syllables according to the Bhatkhande notation (as in Figure~\ref{fig:Figure 1}). The position mapping of the realised syllable onsets to the corresponding canonical syllable was implemented following the method proposed in previous work \cite{WIMAGA}. We observe from Figure~\ref{fig:Figure 2} how the realised onsets lag the canonical locations most of the time.

Finally the vocal pitch is extracted at 10 ms intervals using an autocorrelation based method for fundamental frequency and voicing \cite{jadoul2018introducing}. Brief pauses and unvoiced regions are linearly interpolated to obtain a continuous pitch contour for each sung syllable region. The pitch contour is converted to cents by normalisation with the known performance's tonic. Eventually, we obtain for each performance in our dataset, the segmented audio of each sung line annotated at the syllable level with syllable name, boundaries and the pitch (cents) at 10 ms intervals. We use these low-level features to define quantities that capture the singing variations across repetitions of a \textit{bandish} line within and across singers. The reference for the comparison is the syllable identity (i.e. its name and metrical location) as defined in the canonical notation as presented in Figure~\ref{fig:Figure 1}. 

\begin{table}[htbp]
    \centering
    \scriptsize
    \begin{tabular}{@{}lcc@{}}  
        \hline
        Raga & Bhimpalasi & Yaman \\
        \hline
        Bandish & Ja Ja Re & Yeri Aali \\
        Tala & Teentaal & Teentaal \\
        Swar & S, R, g, m, P, D, n, S & S, R, G, M, P, D, N, S \\
        \# Concerts & 15 & 13 \\
        \# Artists & 15 & 12 \\
        \# Repetitions (L1, L2, L3, L4) & 167, 39, 47, 47 & 94, 32, 35, 23 \\
        Matra per min range & 138--200 & 111--203 \\
        \hline
    \end{tabular}
    \caption{Summary of our dataset of concert recordings across \textit{ragas}, \textit{bandish}, and artists. \textit{Swar} notation details are in the supplementary.}
    \label{tab:datasummary}
\end{table}





\section{Measuring Expressiveness}\label{sec:observations}
We wish to quantify and compare the variability observed in the acoustic realisation of a given syllable, from a specific line of the \textit{bandish}, across (i) repeated utterances within an artist’s performance, and (ii) utterances of the same syllable across different performances/artists. The acoustic parameters that we detect are: (i) the onset time of the syllable, (ii) the syllable duration (as the time interval between the current syllable's onset and either the onset of the next syllable or the start of the following silence segment, whichever occurs first. and (iii) the pitch contour shape across the syllable interval. We illustrate the process by providing examples of the processing and analyses of the audio rendering of a chosen line by one artist.  

\subsection{Timing expression}\label{subsec:body}

The deviation of the detected onset from its reference assigned beat index in the canonical notation gives us an estimate of the lag/lead of the singer for the syllable in question. We represent the deviation in terms of fraction of the local beat interval; this normalization facilitates the comparison across instances and concerts. We can view the thus measured timing offsets as evidence of expressive timing, especially if this quantity shows variability across repetitions of the syllable within the concert.

\begin{figure}[ht]
    \centering
    \includegraphics[width=1\linewidth]{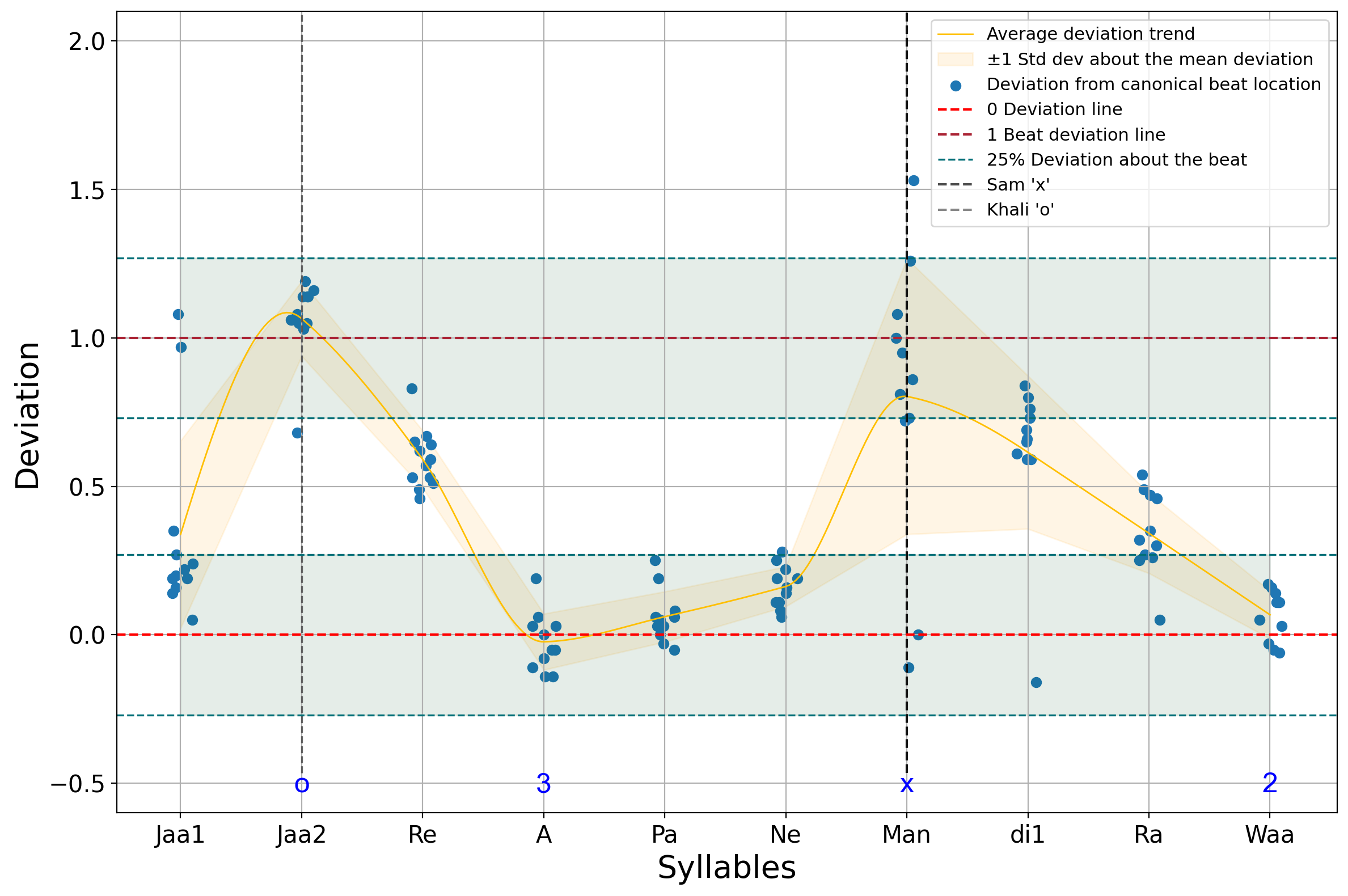}
    \caption{Deviation of the sung syllable onsets from the canonical locations measured in the units of beat duration for \textit{Ja Ja Re} Line 1 by ABD for multiple repetitions of the line.}
    \label{fig:Figure 3}
\end{figure}

Figure~\ref{fig:Figure 3} captures the onsets of syllables in \textit{bandish} 1, Line 1 as rendered by singer ABD. The syllable names are shown with their canonical \textit{matra} locations at the bottom. We note that some syllables occupy 2 beats and others 1 beat in the canonical form. We observe, for example, that "Jaa2" and "Man" (both 2-beat syllables) show variations with a mean lag of about one beat. "Man" instances however are much more dispersed. While a uniform offset could potentially indicate a structural difference between the artist's version of the \textit{bandish} and that of the Bhatkhande book, a high standard deviation (like in "Man") points to the artist's in-the-moment expressive variations. On the other hand, the 3 syllables preceding "Man" show near-zero offsets across repetitions.



\subsection{Pitch expression}


Analysing the rendered song for pitch-based expressiveness is a rather involved task. There are many ways in which the artist injects expressiveness in their performance via pitch variation. As we want to analyse the different ways of realising the same line of a \textit{bandish}, we are interested in the variability in the pitch contour (PC) shape of a syllable across the multiple repetitions of that line. A greater diversity of PC shapes can then be interpreted as higher expressiveness in an artist's repetitions of the line.  

\begin{figure}[ht]
    \centering
    \includegraphics[width=1\linewidth]{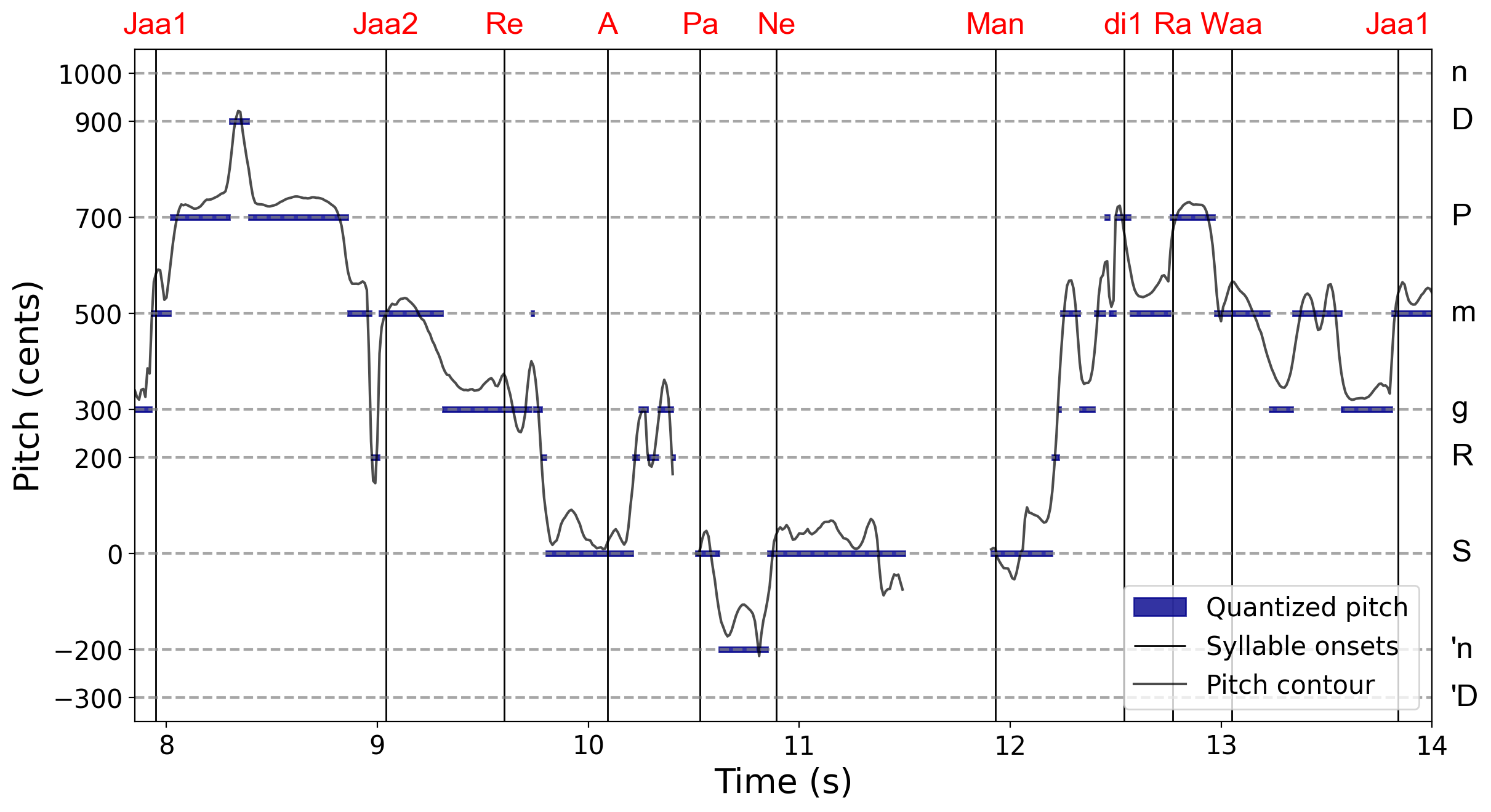}
    \caption{PC quantisation to the nearest \textit{raga} note for one instance of \textit{Ja Ja Re} \textit{bandish} line 1 rendered by artist ABD}
    \label{fig:figure_pitch_quant}
\end{figure}

\begin{figure}[ht]
    \centering
    \includegraphics[width=1\linewidth]{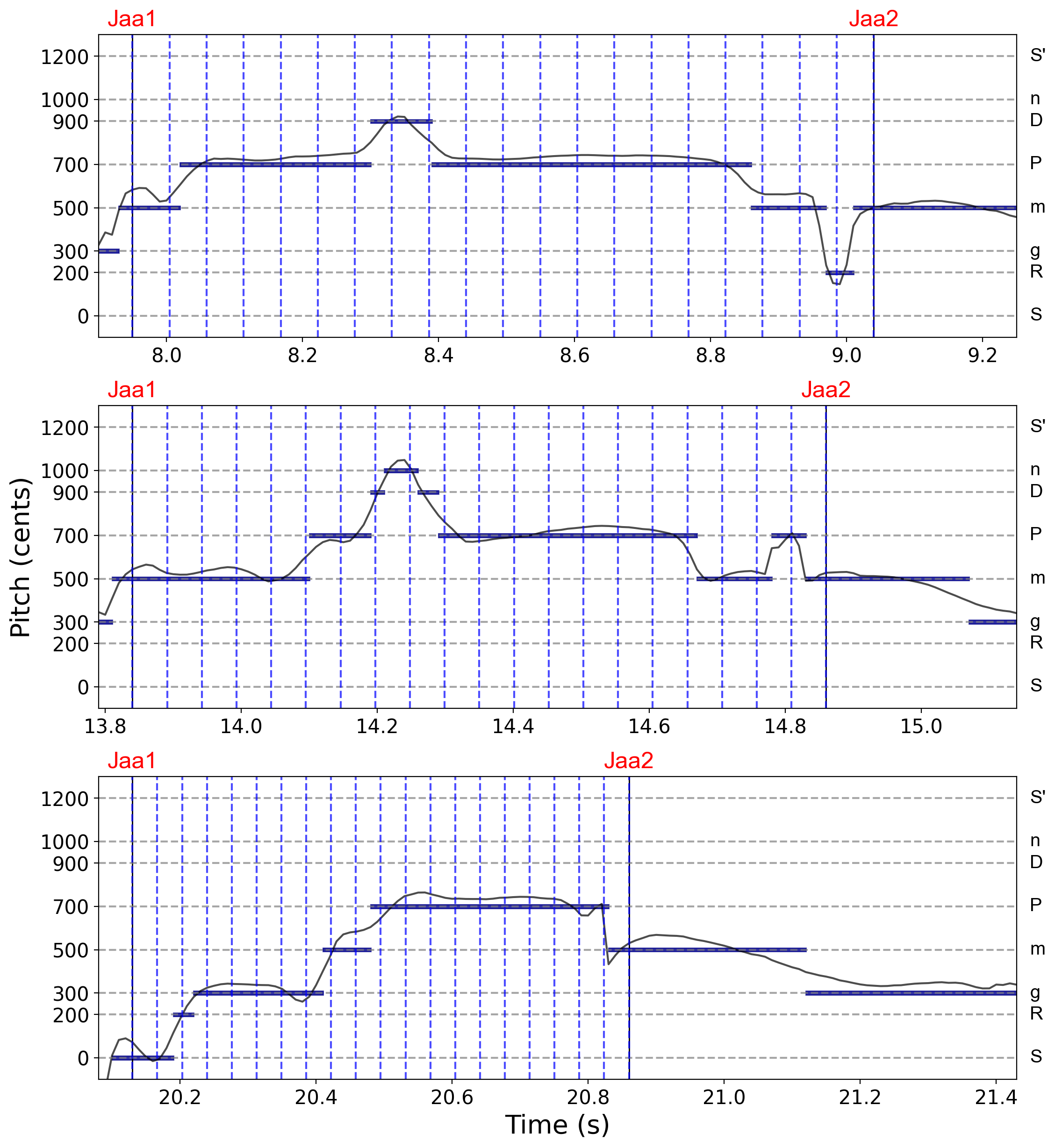}
    \caption{Three distinct renditions of the syllable "Jaa1" by artist ABD, each represented by a fixed number of uniform time intervals. Each interval is mapped based on its modal pitch to the nearest \textit{raga} note, giving us the PAA string representation for the syllable's pitch shape. Figure~\ref{fig:Figure 6} describes this process and the PAA strings for the above PCs.} 
    \label{fig:Figure 5}
\end{figure}

For each syllable, the associated PC spans the duration of that syllable; hence, this is not a fixed-length time series. We represent this variable dimension PC for a given syllable with a lower and fixed-dimensional vector. We first implement a piece-wise aggregate approximation (PAA) over the syllable PCs. PAA is a time-series representation that has been used widely in data-mining tasks \cite{Keogh}. This is implemented as follows. 

\begin{figure}[ht]
    \centering
    \includegraphics[width=1\linewidth]{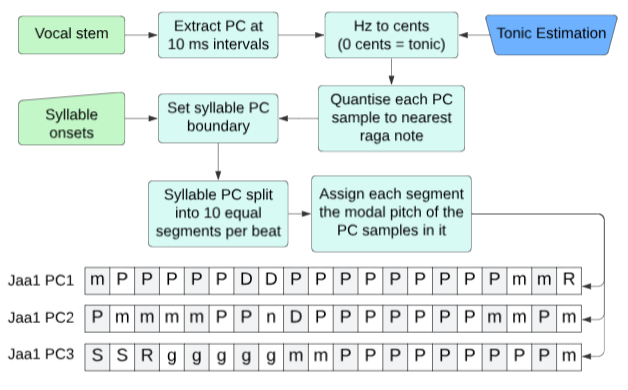}
    \caption{Processing pipeline for generating PAA string representations for the pitch contour for different repetitions of a syllable. The strings correspond to the PCs of the syllable repetitions given in Figure~\ref{fig:Figure 5} spanning 2 beats. }
    \label{fig:Figure 6}
\end{figure}

The syllable PC values are each first quantised to the nearest \textit{raga} \textit{swar} (note), Figure~\ref{fig:figure_pitch_quant} shows this process. Next, the quantised PC values for a syllable across repetitions, is aggregated by dividing the quantised PC into a fixed number of uniform intervals and assigning the mode of the values to each interval. The number of fixed intervals is set empirically to 10 intervals per beats allotted to the syllable (treating the syllable extensions indicated by '-' in Figure~\ref{fig:Figure 1} to be a part of the previous syllable, thereby adding to its allotted beats). The choice of 10 equal segments per beat interval is based on the tempo range of our dataset (110-200 BPM or 300 ms to 545 ms per beat) and sampling period (10 ms/sample) of the pitch contour. This results in segments, each represented by a short sequence of samples of the pitch contour. This balance allows capturing dynamic pitch fluctuations which are prevalent in Hindustani classical music (HCM), without over-quantisation. Now, each PAA interval within a syllable is assigned a discrete symbol, where the symbols are drawn from a suitable alphabet which comprises notes (\textit{swar}) across the relevant octave ranges. The resulting string of note values (one per PAA interval) then represents coarsely the realised pitch shape of the syllable. Figure~\ref{fig:Figure 6} shows this process. The 3 string sequences in Figure~\ref{fig:Figure 6} correspond to the 3 PCs in Figure~\ref{fig:Figure 5}. This type of aggregation presents a tradeoff of generality vs specificity.

\begin{figure}[ht]
    \centering
    \includegraphics[width=1\linewidth]{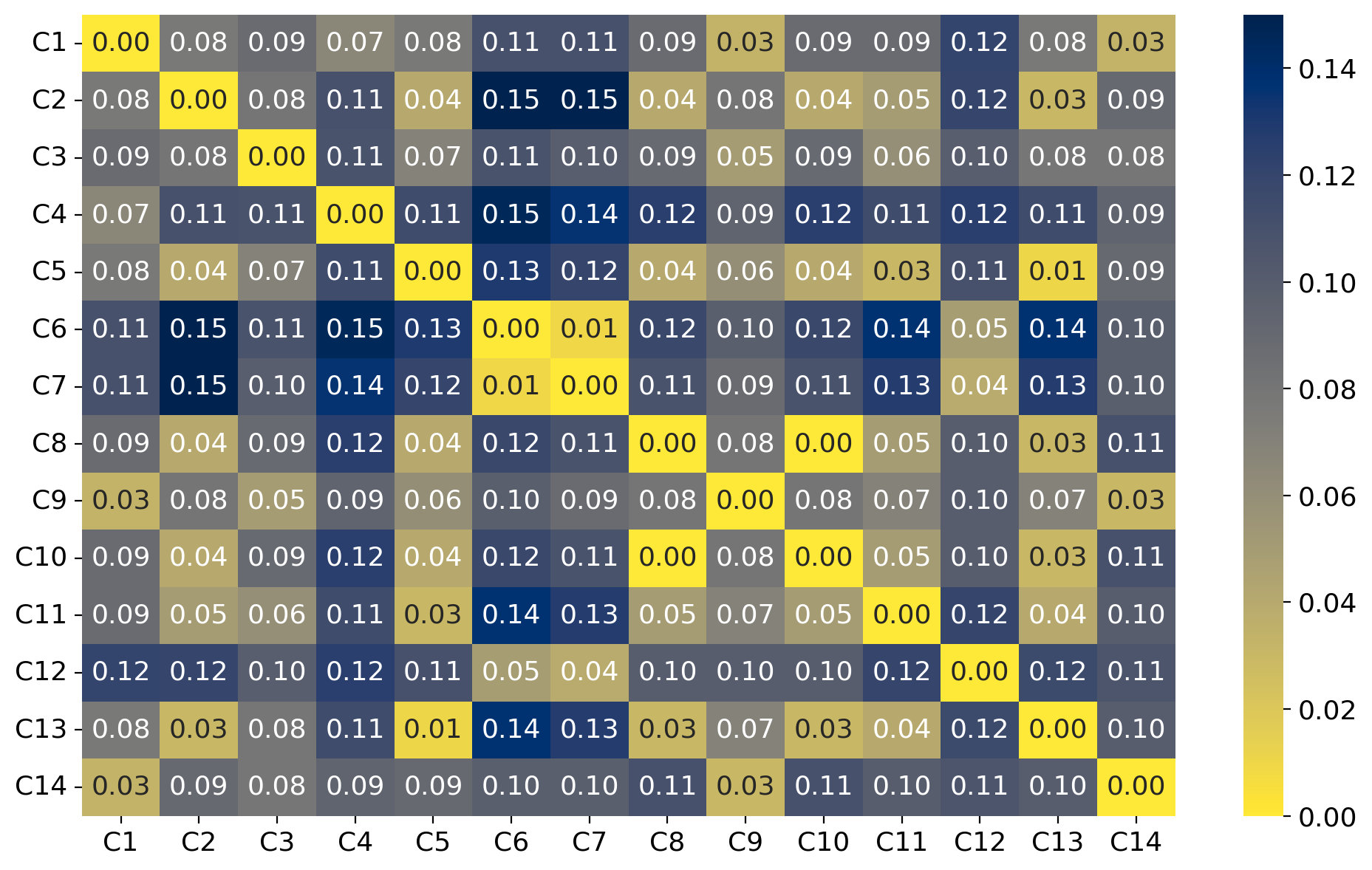}
    \caption{Heat map showing NLSS for each pair of "Jaa1" syllable PCs drawn from the set of repetitions of line 1 of \textit{bandish} \textit{Ja Ja Re} rendered by artist ABD }
    \label{fig:Figure 7}
\end{figure}

Like in the case of syllable timing, we are interested in the variation, if any, in pitch shape of a given syllable across repetitions. We achieve this by computing the similarity of the syllable PCs for pairs drawn from the set of repetitions in a single concert. The Levenshtein edit distance \cite{levens2004measuring} between the PAA strings provides us with the number of note substitutions. We evaluate the Normalised Levenshtein Substitution Score (NLSS) for each pair as a measure of the dissimilarity. Figure~\ref{fig:Figure 7} shows a matrix representation (heat map) of NLSS values for a chosen syllable as rendered by one artist across 14 repetitions.

\section{Observations and Discussion}

We are interested in the within-artist variation for a given \textit{bandish} line and its syllables. This can facilitate potentially valuable insights about (i) the preferred locations (in terms of chosen syllables) for expressive gestures by individual artists and across artists and (ii) the extent and nature of expressive gestures for a given artist. Due to space limitations, we present the analysis results for line 1 of one \textit{bandish}, with the other \textit{bandish} presented in the supplementary.


Figure~\ref{fig:tableTiming} presents for each artist and syllable, the standard deviation (s.d.) of the timing deviation (as captured in the example for artist ABD in {Figure~\ref{fig:Figure 3}}). The s.d. helps us focus on the \textit{variability} of offsets rather than on actual offset values (which might be attributed to structural differences between the artist's version and Bhatkhande's version of the \textit{bandish}, rather than expression-related).

In Figure~\ref{fig:tableTiming}, we note the dominance of the first 3 syllables for most artists. The full range of behaviours, however, includes IN at one end with minimal variations to DG and RK, who introduce new variations on practically all syllables. That IN does not exercise any flexibility is not surprising given that his performances were explicitly created to closely follow the prescribed Bhatkhande notation, as discussed here \cite{Scroll}. RK, on the other hand, is considered a virtuoso musician. 

Aggregating across the rows, we obtain the per syllable behaviour across artists in Figure~\ref{fig:Boxplot_timingDev}. The mean values show that the first 3 syllables carry the most expressive timing, with "Jaa2" also showing the most spread across artists (consistent with Figure~\ref{fig:tableTiming}). We see, for example, that PT shows a large range in per syllable SD, again agreeing with Figure~\ref{fig:tableTiming}. In the case of the artist ABD, we can observe that the temporal deviation is spread relatively evenly, while peaking for a particular syllable "Man", which is the downbeat. This can be easily appreciated in listening to the audio, which can be accessed in the supplementary material. 


\begin{figure}
    \centering
    \includegraphics[width=1\linewidth]{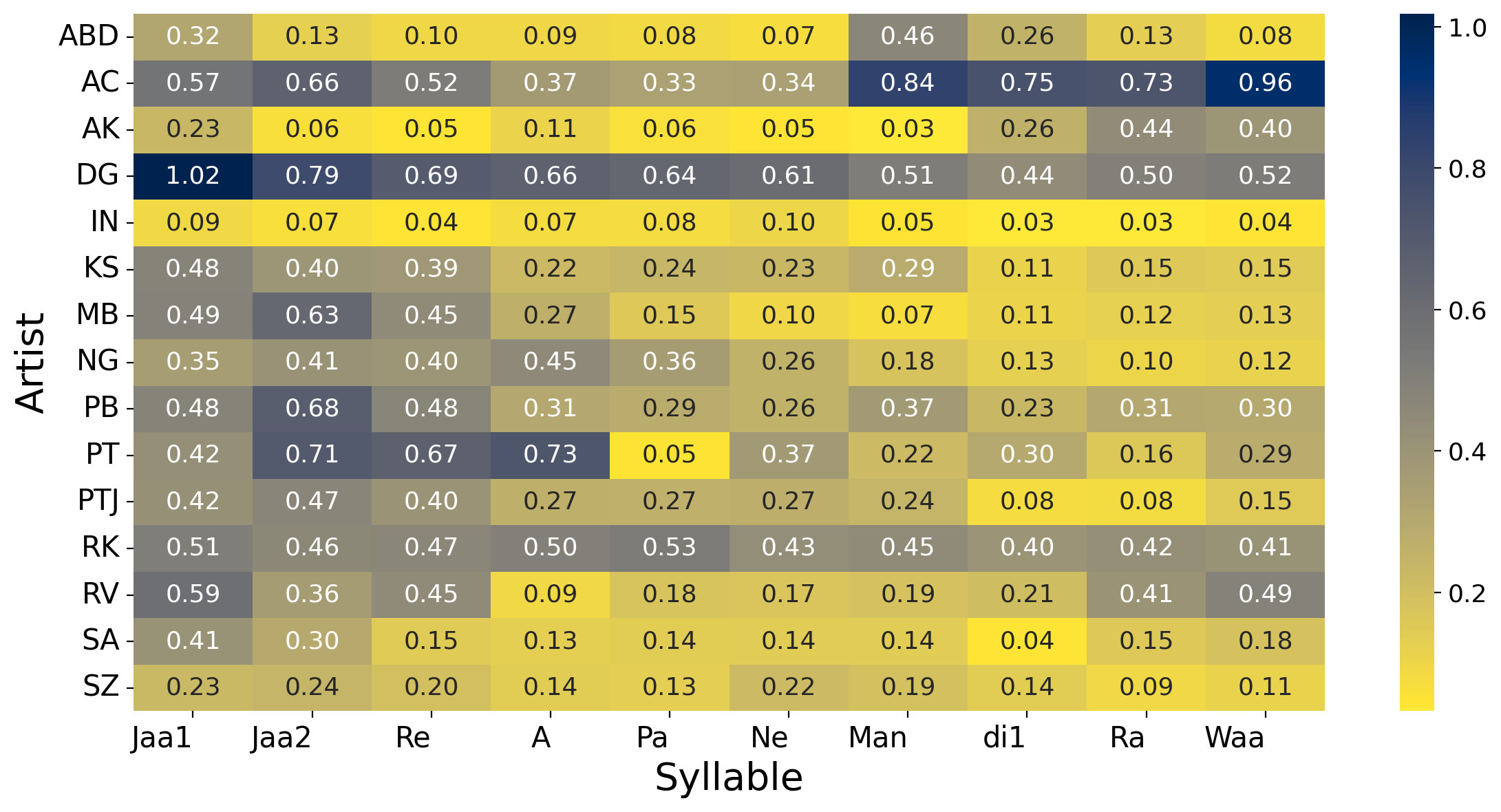}
    \caption{Standard deviation of the distribution of the fractional timing deviation for every syllable over multiple repetitions in one rendition, across artists.}
    \label{fig:tableTiming}
\end{figure}

\begin{figure}
    \centering
    \includegraphics[width=1\linewidth]{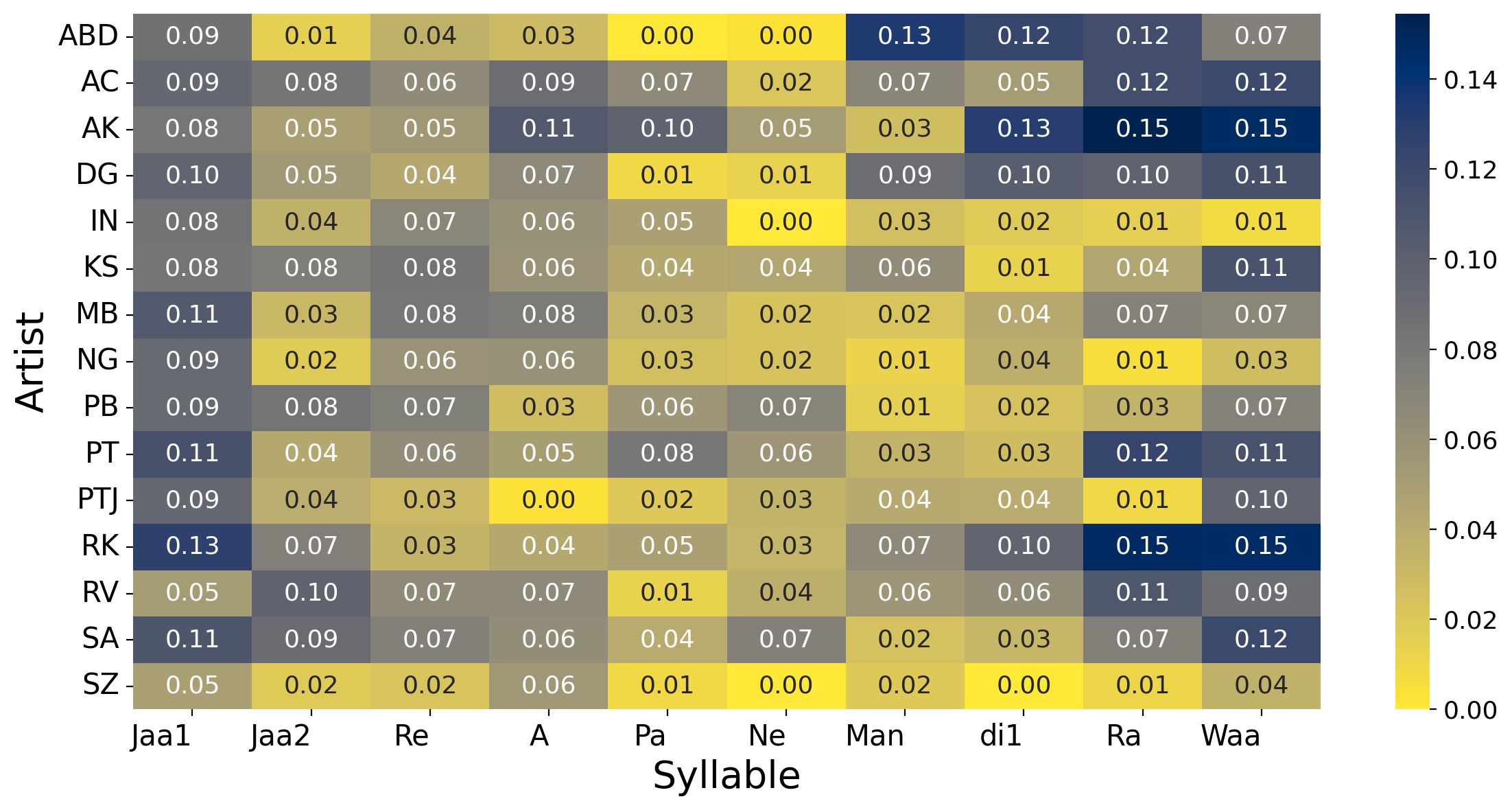}
    \caption{Mean NLSS over all pairs of repetitions of each syllable in line 1 of \textit{bandish} \textit{Ja Ja Re}, computed per artist. serving as a dissimilarity measure across different repetitions of a syllable by an artist.}
    \label{fig:tablePitch}
\end{figure}

To assess pitch variability, we calculate the average number of pitch substitutions by taking the mean of the NLSS for all pairs of repetitions per artist and per syllable, (normalised by the string length), calculated across all possible pairs of the given syllable utterances within a concert. Figure~\ref{fig:tablePitch} shows the values per artist and per syllable of Line 1. We can observe in Figure~\ref{fig:tableTiming} and Figure~\ref{fig:tablePitch} that both the pitch and temporal variation across repetitions and across different artists is more prominent at the beginning of the line. Near the end of the line, the pitch variation increases, which can be attributed to the emphasis on a semantically important word- "Mandirava".

\begin{figure}
    \centering
    \includegraphics[width=1\linewidth]{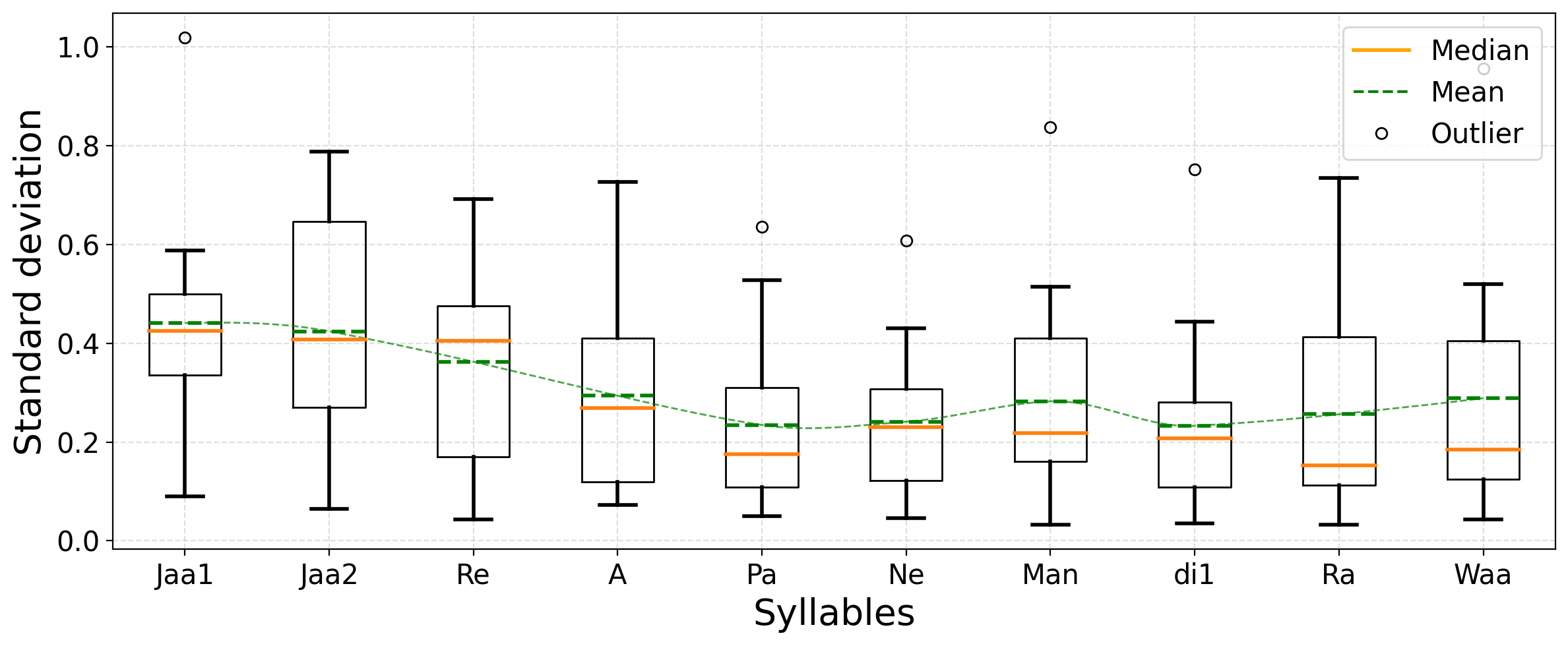}
    \caption{Box-plot of s.d. of timing deviation parameter per syllable of \textit{Ja Ja Re} Line 1 as aggregated across concerts and artists. The \textit{tala}-cycle ends at the syllable "Ne", and a new cycle starts from downbeat at the syllable "Man".}
    \label{fig:Boxplot_timingDev}
\end{figure}

The syllables "Pa" and "Ne" are near the \textit{tala}-cycle boundary, and we observe a minimum pitch and temporal variation; this is consistent with the observations of \cite{WIMAGA}. As the performer heads towards the \textit{tala}-cycle boundary, their overall tendency of variation and improvisation decreases, aiming rather towards resolving the melodic and temporal expression they have come up with in the particular \textit{tala}-cycle.

\begin{figure}
    \centering
    \includegraphics[width=1\linewidth]{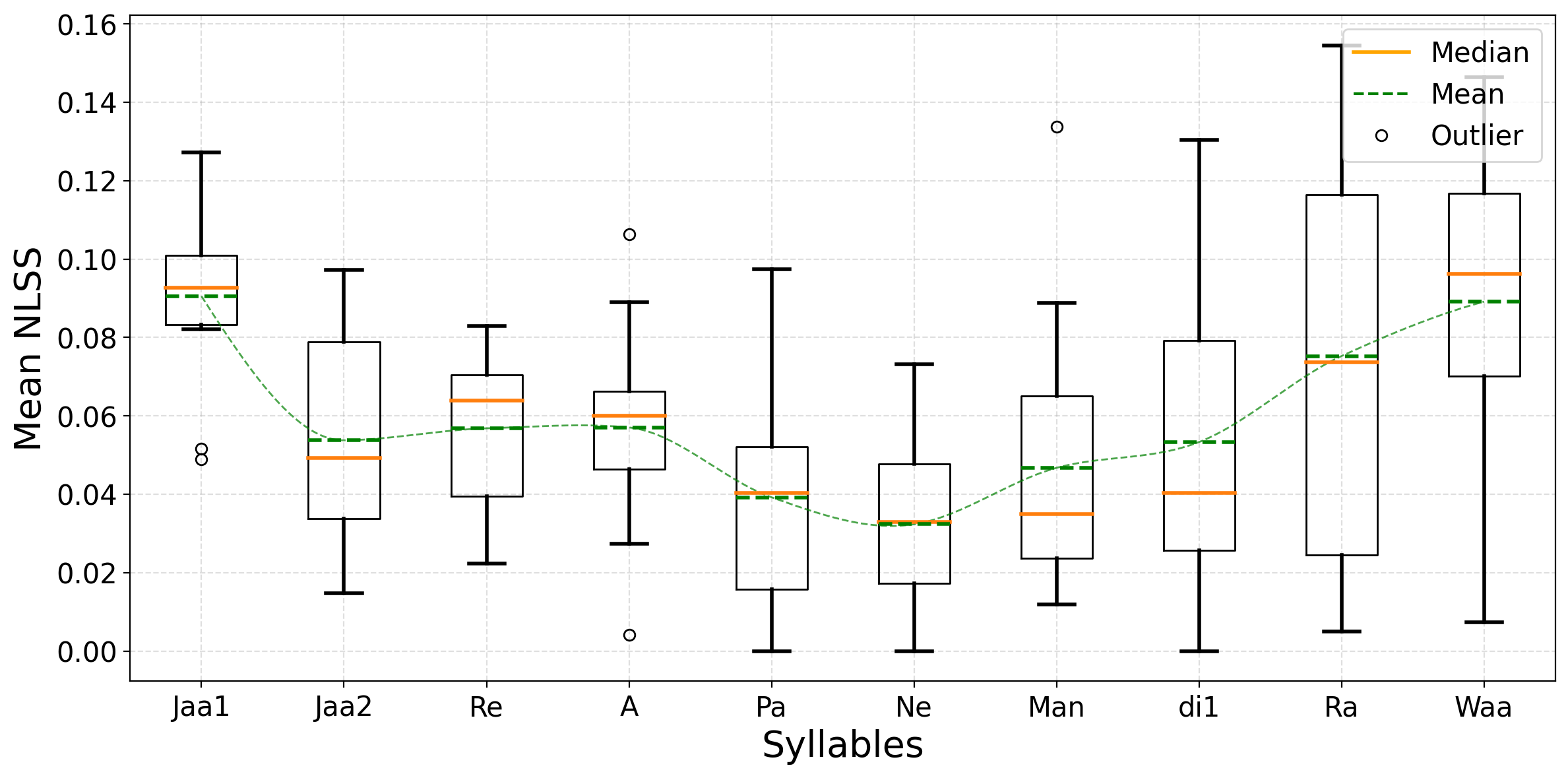}
    \caption{Box-plot of the mean of the averaged NLSS per syllable of \textit{Ja Ja Re} Line 1 as aggregated across concerts and artists.}
    \label{fig:Boxplot_pitchDev}
\end{figure}

Aggregating across rows of Figure~\ref{fig:tablePitch}, we obtain the per syllable behaviour over all artists in Figure~\ref{fig:Boxplot_pitchDev}. The dotted line joining the means indicates that the pitch variation decreases as one approaches the \textit{tala}-cycle boundary at the syllable "Ne", while the the pitch expressiveness is higher in the start and end of the line, which falls on the 7th and 5th beats of the \textit{tala}-cycle respectively, which is far from the cycle boundary, hence providing more scope for timing and pitch expressiveness. We see, for example, that AC and AK show high pitch variation across the repetitions of the same line.


Hierarchical clustering of all the pairs from the set of repetitions of a syllable by an artist provides us with information about the variation clusters. A threshold can be defined that decides if a variation belongs to a cluster or not. More diverse variations would indicate more number of clusters, indicating higher expressiveness. Dendrograms are excellent for visualising such clusters. The realisation of the previous syllable has an influence on which cluster the following syllable variation would belong to.


Comparing Figure~\ref{fig:tableTiming} and Figure~\ref{fig:tablePitch}, we note that expressive gestures that utilise pitch are not necessarily at the same locations that exhibit timing deviation in terms of the preferred syllable. It is rather interesting to look at the least amount of expressiveness in both pitch and timing lie with the syllables that are near the \textit{tala}-cycle boundary. An analysis at the individual audio level would provide a more accurate picture of the correlations, if any, and is left to future work.

Our observations, reported here on the Line 1 of one \textit{bandish}, largely hold with the second \textit{bandish}. The underlying reasons for the choice of specific syllables exhibiting larger variability are similar to those discussed by Morris\cite{Morris-AD}. These include larger variations at line or phrase ending syllables due to the effect of previous and next contexts, and the choice of syllables belonging to more emotionally loaded words in the lyrics.

\section{Conclusion}
In this paper, we articulated the problem of modeling expressive variations in the context of performance of Hindustani traditional compositions by established artists of the genre. A well-known \textit{bandish} in the chosen \textit{raga} is always sung at the beginning of a concert with multiple repetitions of the lines, marked by variations in the location and type of the expressive gestures. Based on our proposed methodology, we showed that it is possible to arrive at systematic patterns across artists by treating the syllables of the lyrics as reference points for a study of the range of variation. This also helped us discuss interesting correlations between the roles of melody and rhythm in expressiveness.

We presented a dataset that was annotated with a combination of manual and automatic tools to obtain a rich repository of distinct realizations of the lines of two popular traditional compositions.  While much further exploration remains possible, this work demonstrates the potential of computational models for improvisation in the context of compositions in the \textit{Khayal} genre. 
This work lays the foundation for generative applications by capturing high-level performance features that reflect an artist’s distinctive style. A preliminary experiment was performed to generate the temporal deviations discussed above, where the distributions formed by all the deviations of a syllable from its canonical location for an artist were used to sample out new points for each syllable. A sine-tone based audio was synthesized from the generated pitch contour. 
The reference (Bhatkhande canonical form) and generated tracks are available in the supplementary. This lets us create infinite possibilities for rendering the same line, while at the same time capturing some hint of the artist's style. Extending this approach to other acoustic dimensions for expression such as timbre and dynamics can enable the generation of classical music that embodies the unique identity of individual artists. 

\section{Acknowledgments}

 We take this opportunity to acknowledge and express our gratitude to all those who supported and guided us during this research work. We are thankful to Madhumitha S., whose thesis work was a crucial base for our project. We also thank Mr. Himanshu Sati, and Mrs. Hemala Ranade for their scholarly musicological insights, which gave direction to our work. 



\bibliography{ISMIRtemplate.bib}

\begin{thebibliography}{10}
\providecommand{\url}[1]{#1}
\csname url@samestyle\endcsname
\providecommand{\newblock}{\relax}
\providecommand{\bibinfo}[2]{#2}
\providecommand{\BIBentrySTDinterwordspacing}{\spaceskip=0pt\relax}
\providecommand{\BIBentryALTinterwordstretchfactor}{4}
\providecommand{\BIBentryALTinterwordspacing}{\spaceskip=\fontdimen2\font plus
\BIBentryALTinterwordstretchfactor\fontdimen3\font minus \fontdimen4\font\relax}
\providecommand{\BIBforeignlanguage}[2]{{%
\expandafter\ifx\csname l@#1\endcsname\relax
\typeout{** WARNING: IEEEtran.bst: No hyphenation pattern has been}%
\typeout{** loaded for the language `#1'. Using the pattern for}%
\typeout{** the default language instead.}%
\else
\language=\csname l@#1\endcsname
\fi
#2}}
\providecommand{\BIBdecl}{\relax}
\BIBdecl

\bibitem{Wade}
B.~Wade, ``Music in {I}ndia: The classical traditions,'' Manohar Press, 2001.

\bibitem{cancino2018computational}
C.~E. Cancino-Chac{\'o}n, M.~Grachten, W.~Goebl, and G.~Widmer, ``Computational models of expressive music performance: A comprehensive and critical review,'' \emph{Frontiers in Digital Humanities}, vol.~5, p.~25, 2018.

\bibitem{juslin2000cue}
P.~N. Juslin, ``Cue utilization in communication of emotion in music performance: Relating performance to perception.'' \emph{Journal of Experimental Psychology: Human perception and performance}, vol.~26, no.~6, p. 1797, 2000.

\bibitem{Morris-AD}
A.~Morris, \emph{Transmission and performance of Khayal compositions in the {G}walior gharana of {I}ndian vocal music.}\hskip 1em plus 0.5em minus 0.4em\relax PhD thesis, Univ. Of London, S.O.A.S., 2004.

\bibitem{Meer2014book}
W.~Van~der Meer, ``Audience response and expressive pitch inflections in a live recording of legendary singer kesar bai kerkar,'' in \emph{Expressiveness in music performance: Empirical approaches across styles and cultures}, D.~Fabian, R.~Timmers, and E.~Schubert, Eds.\hskip 1em plus 0.5em minus 0.4em\relax Oxford University Press (UK), 2014, pp. 170--184.

\bibitem{sankaran2015automatic}
S.~Sankaran, P.~V.~K. Sekhar, and A.~M. Hema, ``Automatic segmentation of composition in carnatic music using time-frequency cfcc templates,'' in \emph{Proceedings of 11th International Symposium on Computer Music Multidisciplinary Research (CMMR)}, 2015.

\bibitem{ganguli2021}
K.~K. Ganguli and P.~Rao, ``A study of variability in raga motifs in performance contexts,'' \emph{Journal of New Music Research}, vol.~50, no.~1, pp. 102--116, 2021.

\bibitem{WIMAGA}
Y.~Bhake and P.~Rao, ``Expressive timing in {H}industani vocal music,'' in \emph{Proc. of ICASSP 2025 Workshop on Indian Music Analysis and Generative Applications (WIMAGA)}, Hyderabad, India, 2025, accessed at:\href{https://arxiv.org/abs/2503.21142}{link}.

\bibitem{srao-hcm}
S.~Rao and P.~Rao, ``An overview of {H}industani music in the context of computational musicology.'' \emph{Journal of New Music Research}, vol.~43, no.~1, 2014.

\bibitem{Bhatkhande}
V.~Bhatkhande, \emph{Kramik Pustaka Malika}.\hskip 1em plus 0.5em minus 0.4em\relax Sangeet Karyalaya Hathras, India, 2013.

\bibitem{povey2011kaldi}
D.~Povey, A.~Ghoshal, G.~Boulianne, L.~Burget, O.~Glembek, N.~Goel, M.~Hannemann, P.~Motlicek, Y.~Qian, P.~Schwarz \emph{et~al.}, ``The kaldi speech recognition toolkit,'' in \emph{IEEE 2011 workshop on automatic speech recognition and understanding}.\hskip 1em plus 0.5em minus 0.4em\relax IEEE Signal Processing Society, 2011.

\bibitem{jadoul2018introducing}
Y.~Jadoul, B.~Thompson, and B.~De~Boer, ``Introducing parselmouth: A {p}ython interface to praat,'' \emph{Journal of Phonetics}, vol.~71, pp. 1--15, 2018.

\bibitem{Keogh}
J.~Lin, E.~Keogh, L.~Wei, and S.~Lonardi, ``Experiencing {SAX}: a novel symbolic representation of time series.'' \emph{Data Mining and knowledge discovery}, vol.~31, no.~B, pp. 107--144, April 2007.

\bibitem{levens2004measuring}
W.~J. Heeringa, ``Measuring dialect pronunciation differences using {L}evenshtein distance,'' 2004.

\bibitem{Scroll}
``2000 classic compositions from {B}hatkhande on cd,'' \url{https://scroll.in/article/726180/}, accessed: 2024-04-10.

\end{thebibliography}

%
%
%
%

\end{document}